\documentclass[conference, twocolumn]{IEEEtran}
\IEEEoverridecommandlockouts

\usepackage{cite}
\usepackage{amsmath,amssymb,amsfonts}
\usepackage{graphicx}
\usepackage{textcomp}
\usepackage{xcolor}
\usepackage{algorithm}
\usepackage{algorithmic}
\usepackage{balance}
\usepackage{dsfont}
\usepackage{subfig}
\usepackage{setspace}
\usepackage{amssymb}
\usepackage{amsfonts}
\usepackage{mathrsfs}
\usepackage{xspace}
\usepackage{bm}
\usepackage{upgreek}

\newcommand{\safemath}[2]{\newcommand{#1}{\ensuremath{#2}\xspace}}

\safemath{\bma}{\mathbf{a}}
\safemath{\bmb}{\mathbf{b}}
\safemath{\bmc}{\mathbf{c}}
\safemath{\bmd}{\mathbf{d}}
\safemath{\bme}{\mathbf{e}}
\safemath{\bmf}{\mathbf{f}}
\safemath{\bmg}{\mathbf{g}}
\safemath{\bmh}{\mathbf{h}}
\safemath{\bmi}{\mathbf{i}}
\safemath{\bmj}{\mathbf{j}}
\safemath{\bmk}{\mathbf{k}}
\safemath{\bml}{\mathbf{l}}
\safemath{\bmm}{\mathbf{m}}
\safemath{\bmn}{\mathbf{n}}
\safemath{\bmo}{\mathbf{o}}
\safemath{\bmp}{\mathbf{p}}
\safemath{\bmq}{\mathbf{q}}
\safemath{\bmr}{\mathbf{r}}
\safemath{\bms}{\mathbf{s}}
\safemath{\bmt}{\mathbf{t}}
\safemath{\bmu}{\mathbf{u}}
\safemath{\bmv}{\mathbf{v}}
\safemath{\bmw}{\mathbf{w}}
\safemath{\bmx}{\mathbf{x}}
\safemath{\bmy}{\mathbf{y}}
\safemath{\bmz}{\mathbf{z}}
\safemath{\bmzero}{\mathbf{0}}
\safemath{\bmone}{\mathbf{1}}

\bmdefine{\biad}{a}
\bmdefine{\bibd}{b}
\bmdefine{\bicd}{c}
\bmdefine{\bidd}{d}
\bmdefine{\bied}{e}
\bmdefine{\bifd}{f}
\bmdefine{\bigd}{g}
\bmdefine{\bihd}{h}
\bmdefine{\biid}{i}
\bmdefine{\bijd}{j}
\bmdefine{\bikd}{k}
\bmdefine{\bild}{l}
\bmdefine{\bimd}{m}
\bmdefine{\bind}{n}
\bmdefine{\biod}{o}
\bmdefine{\bipd}{p}
\bmdefine{\biqd}{q}
\bmdefine{\bird}{r}
\bmdefine{\bisd}{s}
\bmdefine{\bitd}{t}
\bmdefine{\biud}{u}
\bmdefine{\bivd}{v}
\bmdefine{\biwd}{w}
\bmdefine{\bixd}{x}
\bmdefine{\biyd}{y}
\bmdefine{\bizd}{z}

\bmdefine{\bixid}{\xi}
\bmdefine{\bilambdad}{\lambda}
\bmdefine{\bimud}{\mu}
\bmdefine{\bithetad}{\theta}
\bmdefine{\biphid}{\phi}
\bmdefine{\bideltad}{\delta}

\safemath{\bmia}{\biad}
\safemath{\bmib}{\bibd}
\safemath{\bmic}{\bicd}
\safemath{\bmid}{\bidd}
\safemath{\bmie}{\bied}
\safemath{\bmif}{\bifd}
\safemath{\bmig}{\bigd}
\safemath{\bmih}{\bihd}
\safemath{\bmii}{\biid}
\safemath{\bmij}{\bijd}
\safemath{\bmik}{\bikd}
\safemath{\bmil}{\bild}
\safemath{\bmim}{\bimd}
\safemath{\bmin}{\bind}
\safemath{\bmio}{\biod}
\safemath{\bmip}{\bipd}
\safemath{\bmiq}{\biqd}
\safemath{\bmir}{\bird}
\safemath{\bmis}{\bisd}
\safemath{\bmit}{\bitd}
\safemath{\bmiu}{\biud}
\safemath{\bmiv}{\bivd}
\safemath{\bmiw}{\biwd}
\safemath{\bmix}{\bixd}
\safemath{\bmiy}{\biyd}
\safemath{\bmiz}{\bizd}

\safemath{\bmxi}{\bixid}
\safemath{\bmlambda}{\bilambdad}
\safemath{\bmmu}{\bimud}
\safemath{\bmtheta}{\bithetad}
\safemath{\bmphi}{\biphid}
\safemath{\bmdelta}{\bideltad}

\safemath{\bA}{\mathbf{A}}
\safemath{\bB}{\mathbf{B}}
\safemath{\bC}{\mathbf{C}}
\safemath{\bD}{\mathbf{D}}
\safemath{\bE}{\mathbf{E}}
\safemath{\bF}{\mathbf{F}}
\safemath{\bG}{\mathbf{G}}
\safemath{\bH}{\mathbf{H}}
\safemath{\bI}{\mathbf{I}}
\safemath{\bJ}{\mathbf{J}}
\safemath{\bK}{\mathbf{K}}
\safemath{\bL}{\mathbf{L}}
\safemath{\bM}{\mathbf{M}}
\safemath{\bN}{\mathbf{N}}
\safemath{\bO}{\mathbf{O}}
\safemath{\bP}{\mathbf{P}}
\safemath{\bQ}{\mathbf{Q}}
\safemath{\bR}{\mathbf{R}}
\safemath{\bS}{\mathbf{S}}
\safemath{\bT}{\mathbf{T}}
\safemath{\bU}{\mathbf{U}}
\safemath{\bV}{\mathbf{V}}
\safemath{\bW}{\mathbf{W}}
\safemath{\bX}{\mathbf{X}}
\safemath{\bY}{\mathbf{Y}}
\safemath{\bZ}{\mathbf{Z}}

\safemath{\bZero}{\mathbf{0}}
\safemath{\bOne}{\mathbf{1}}
\safemath{\bDelta}{\mathbf{\Delta}}
\safemath{\bLambda}{\mathbf{\UpLambda}}
\safemath{\bPhi}{\mathbf{\Upphi}}
\safemath{\bSigma}{\mathbf{\Upsigma}}
\safemath{\bOmega}{\mathbf{\Upomega}}
\safemath{\bTheta}{\mathbf{\Uptheta}}

\bmdefine{\biAd}{A}
\bmdefine{\biBd}{B}
\bmdefine{\biCd}{C}
\bmdefine{\biDd}{D}
\bmdefine{\biEd}{E}
\bmdefine{\biFd}{F}
\bmdefine{\biGd}{G}
\bmdefine{\biHd}{H}
\bmdefine{\biId}{I}
\bmdefine{\biJd}{J}
\bmdefine{\biKd}{K}
\bmdefine{\biLd}{L}
\bmdefine{\biMd}{M}
\bmdefine{\biOd}{N}
\bmdefine{\biPd}{O}
\bmdefine{\biQd}{P}
\bmdefine{\biRd}{R}
\bmdefine{\biSd}{S}
\bmdefine{\biTd}{T}
\bmdefine{\biUd}{U}
\bmdefine{\biVd}{V}
\bmdefine{\biWd}{W}
\bmdefine{\biXd}{X}
\bmdefine{\biYd}{Y}
\bmdefine{\biZd}{Z}

\bmdefine{\biDelta}{\Delta}
\bmdefine{\biLambda}{\Lambda}
\bmdefine{\biPhi}{\Phi}
\bmdefine{\biSigma}{\Sigma}
\bmdefine{\biOmega}{\Omega}
\bmdefine{\biTheta}{\Theta}

\safemath{\bimA}{\biAd}
\safemath{\bimB}{\biBd}
\safemath{\bimC}{\biCd}
\safemath{\bimD}{\biDd}
\safemath{\bimE}{\biEd}
\safemath{\bimF}{\biFd}
\safemath{\bimG}{\biGd}
\safemath{\bimH}{\biHd}
\safemath{\bimI}{\biId}
\safemath{\bimJ}{\biJd}
\safemath{\bimK}{\biKd}
\safemath{\bimL}{\biLd}
\safemath{\bimM}{\biMd}
\safemath{\bimN}{\biNd}
\safemath{\bimO}{\biOd}
\safemath{\bimP}{\biPd}
\safemath{\bimQ}{\biQd}
\safemath{\bimR}{\biRd}
\safemath{\bimS}{\biSd}
\safemath{\bimT}{\biTd}
\safemath{\bimU}{\biUd}
\safemath{\bimV}{\biVd}
\safemath{\bimW}{\biWd}
\safemath{\bimX}{\biXd}
\safemath{\bimY}{\biYd}
\safemath{\bimZ}{\biZd}

\safemath{\bimDelta}{\biDelta}
\safemath{\bimLambda}{\biLambda}
\safemath{\bimPhi}{\biPhi}
\safemath{\bimSigma}{\biSigma}
\safemath{\bimOmega}{\biOmega}
\safemath{\bimTheta}{\biTheta}

\safemath{\setA}{\mathcal{A}}
\safemath{\setB}{\mathcal{B}}
\safemath{\setC}{\mathcal{C}}
\safemath{\setD}{\mathcal{D}}
\safemath{\setE}{\mathcal{E}}
\safemath{\setF}{\mathcal{F}}
\safemath{\setG}{\mathcal{G}}
\safemath{\setH}{\mathcal{H}}
\safemath{\setI}{\mathcal{I}}
\safemath{\setJ}{\mathcal{J}}
\safemath{\setK}{\mathcal{K}}
\safemath{\setL}{\mathcal{L}}
\safemath{\setM}{\mathcal{M}}
\safemath{\setN}{\mathcal{N}}
\safemath{\setO}{\mathcal{O}}
\safemath{\setP}{\mathcal{P}}
\safemath{\setQ}{\mathcal{Q}}
\safemath{\setR}{\mathcal{R}}
\safemath{\setS}{\mathcal{S}}
\safemath{\setT}{\mathcal{T}}
\safemath{\setU}{\mathcal{U}}
\safemath{\setV}{\mathcal{V}}
\safemath{\setW}{\mathcal{W}}
\safemath{\setX}{\mathcal{X}}
\safemath{\setY}{\mathcal{Y}}
\safemath{\setZ}{\mathcal{Z}}
\safemath{\emptySet}{\varnothing}

\safemath{\colA}{\mathscr{A}}
\safemath{\colB}{\mathscr{B}}
\safemath{\colC}{\mathscr{C}}
\safemath{\colD}{\mathscr{D}}
\safemath{\colE}{\mathscr{E}}
\safemath{\colF}{\mathscr{F}}
\safemath{\colG}{\mathscr{G}}
\safemath{\colH}{\mathscr{H}}
\safemath{\colI}{\mathscr{I}}
\safemath{\colJ}{\mathscr{J}}
\safemath{\colK}{\mathscr{K}}
\safemath{\colL}{\mathscr{L}}
\safemath{\colM}{\mathscr{M}}
\safemath{\colN}{\mathscr{N}}
\safemath{\colO}{\mathscr{O}}
\safemath{\colP}{\mathscr{P}}
\safemath{\colQ}{\mathscr{Q}}
\safemath{\colR}{\mathscr{R}}
\safemath{\colS}{\mathscr{S}}
\safemath{\colT}{\mathscr{T}}
\safemath{\colU}{\mathscr{U}}
\safemath{\colV}{\mathscr{V}}
\safemath{\colW}{\mathscr{W}}
\safemath{\colX}{\mathscr{X}}
\safemath{\colY}{\mathscr{Y}}
\safemath{\colZ}{\mathscr{Z}}

\safemath{\opA}{\mathbb{A}}
\safemath{\opB}{\mathbb{B}}
\safemath{\opC}{\mathbb{C}}
\safemath{\opD}{\mathbb{D}}
\safemath{\opE}{\mathbb{E}}
\safemath{\opF}{\mathbb{F}}
\safemath{\opG}{\mathbb{G}}
\safemath{\opH}{\mathbb{H}}
\safemath{\opI}{\mathbb{I}}
\safemath{\opJ}{\mathbb{J}}
\safemath{\opK}{\mathbb{K}}
\safemath{\opL}{\mathbb{L}}
\safemath{\opM}{\mathbb{M}}
\safemath{\opN}{\mathbb{N}}
\safemath{\opO}{\mathbb{O}}
\safemath{\opP}{\mathbb{P}}
\safemath{\opQ}{\mathbb{Q}}
\safemath{\opR}{\mathbb{R}}
\safemath{\opS}{\mathbb{S}}
\safemath{\opT}{\mathbb{T}}
\safemath{\opU}{\mathbb{U}}
\safemath{\opV}{\mathbb{V}}
\safemath{\opW}{\mathbb{W}}
\safemath{\opX}{\mathbb{X}}
\safemath{\opY}{\mathbb{Y}}
\safemath{\opZ}{\mathbb{Z}}
\safemath{\opZero}{\mathbb{O}}
\safemath{\identityop}{\opI}

\safemath{\veca}{\bma}
\safemath{\vecb}{\bmb}
\safemath{\vecc}{\bmc}
\safemath{\vecd}{\bmd}
\safemath{\vece}{\bme}
\safemath{\vecf}{\bmf}
\safemath{\vecg}{\bmg}
\safemath{\vech}{\bmh}
\safemath{\veci}{\bmi}
\safemath{\vecj}{\bmj}
\safemath{\veck}{\bmk}
\safemath{\vecl}{\bml}
\safemath{\vecm}{\bmm}
\safemath{\vecn}{\bmn}
\safemath{\veco}{\bmo}
\safemath{\vecp}{\bmp}
\safemath{\vecq}{\bmq}
\safemath{\vecr}{\bmr}
\safemath{\vecs}{\bms}
\safemath{\vect}{\bmt}
\safemath{\vecu}{\bmu}
\safemath{\vecv}{\bmv}
\safemath{\vecw}{\bmw}
\safemath{\vecx}{\bmx}
\safemath{\vecy}{\bmy}
\safemath{\vecz}{\bmz}

\safemath{\veczero}{\bmzero}
\safemath{\vecone}{\bmone}
\safemath{\vecxi}{\bmxi}
\safemath{\veclambda}{\bmlambda}
\safemath{\vecmu}{\bmmu}
\safemath{\vectheta}{\bmtheta}
\safemath{\vecphi}{\bmphi}
\safemath{\vecdelta}{\bmdelta}

\safemath{\matA}{\bA}
\safemath{\matB}{\bB}
\safemath{\matC}{\bC}
\safemath{\matD}{\bD}
\safemath{\matE}{\bE}
\safemath{\matF}{\bF}
\safemath{\matG}{\bG}
\safemath{\matH}{\bH}
\safemath{\matI}{\bI}
\safemath{\matJ}{\bJ}
\safemath{\matK}{\bK}
\safemath{\matL}{\bL}
\safemath{\matM}{\bM}
\safemath{\matN}{\bN}
\safemath{\matO}{\bO}
\safemath{\matP}{\bP}
\safemath{\matQ}{\bQ}
\safemath{\matR}{\bR}
\safemath{\matS}{\bS}
\safemath{\matT}{\bT}
\safemath{\matU}{\bU}
\safemath{\matV}{\bV}
\safemath{\matW}{\bW}
\safemath{\matX}{\bX}
\safemath{\matY}{\bY}
\safemath{\matZ}{\bZ}
\safemath{\matzero}{\bmzero}

\safemath{\matDelta}{\bDelta}
\safemath{\matLambda}{\bLambda}
\safemath{\matPhi}{\bPhi}
\safemath{\matSigma}{\bSigma}
\safemath{\matOmega}{\bOmega}
\safemath{\matTheta}{\bTheta}

\safemath{\matidentity}{\matI}
\safemath{\matone}{\matO}

\safemath{\rnda}{A}
\safemath{\rndb}{B}
\safemath{\rndc}{C}
\safemath{\rndd}{D}
\safemath{\rnde}{E}
\safemath{\rndf}{F}
\safemath{\rndg}{G}
\safemath{\rndh}{H}
\safemath{\rndi}{I}
\safemath{\rndj}{J}
\safemath{\rndk}{K}
\safemath{\rndl}{L}
\safemath{\rndm}{M}
\safemath{\rndn}{N}
\safemath{\rndo}{O}
\safemath{\rndp}{P}
\safemath{\rndq}{Q}
\safemath{\rndr}{R}
\safemath{\rnds}{S}
\safemath{\rndt}{T}
\safemath{\rndu}{U}
\safemath{\rndv}{V}
\safemath{\rndw}{W}
\safemath{\rndx}{X}
\safemath{\rndy}{Y}
\safemath{\rndz}{Z}

\safemath{\rveca}{\bimA}
\safemath{\rvecb}{\bimB}
\safemath{\rvecc}{\bimC}
\safemath{\rvecd}{\bimD}
\safemath{\rvece}{\bimE}
\safemath{\rvecf}{\bimF}
\safemath{\rvecg}{\bimG}
\safemath{\rvech}{\bimH}
\safemath{\rveci}{\bimI}
\safemath{\rvecj}{\bimJ}
\safemath{\rveck}{\bimK}
\safemath{\rvecl}{\bimL}
\safemath{\rvecm}{\bimM}
\safemath{\rvecn}{\bimN}
\safemath{\rveco}{\bomO}
\safemath{\rvecp}{\bimP}
\safemath{\rvecq}{\bimQ}
\safemath{\rvecr}{\bimR}
\safemath{\rvecs}{\bimS}
\safemath{\rvect}{\bimT}
\safemath{\rvecu}{\bimU}
\safemath{\rvecv}{\bimV}
\safemath{\rvecw}{\bimW}
\safemath{\rvecx}{\bimX}
\safemath{\rvecy}{\bimY}
\safemath{\rvecz}{\bimZ}

\safemath{\rvecxi}{\bmxi}
\safemath{\rveclambda}{\bmlambda}
\safemath{\rvecmu}{\bmmu}
\safemath{\rvectheta}{\bmtheta}
\safemath{\rvecphi}{\bmphi}

\safemath{\rmatA}{\bimA}
\safemath{\rmatB}{\bimB}
\safemath{\rmatC}{\bimC}
\safemath{\rmatD}{\bimD}
\safemath{\rmatE}{\bimE}
\safemath{\rmatF}{\bimF}
\safemath{\rmatG}{\bimG}
\safemath{\rmatH}{\bimH}
\safemath{\rmatI}{\bimI}
\safemath{\rmatJ}{\bimJ}
\safemath{\rmatK}{\bimK}
\safemath{\rmatL}{\bimL}
\safemath{\rmatM}{\bimM}
\safemath{\rmatN}{\bimN}
\safemath{\rmatO}{\bimO}
\safemath{\rmatP}{\bimP}
\safemath{\rmatQ}{\bimQ}
\safemath{\rmatR}{\bimR}
\safemath{\rmatS}{\bimS}
\safemath{\rmatT}{\bimT}
\safemath{\rmatU}{\bimU}
\safemath{\rmatV}{\bimV}
\safemath{\rmatW}{\bimW}
\safemath{\rmatX}{\bimX}
\safemath{\rmatY}{\bimY}
\safemath{\rmatZ}{\bimZ}

\safemath{\rmatDelta}{\bimDelta}
\safemath{\rmatLambda}{\bimLambda}
\safemath{\rmatPhi}{\bimPhi}
\safemath{\rmatSigma}{\bimSigma}
\safemath{\rmatOmega}{\bimOmega}
\safemath{\rmatTheta}{\bimTheta}

\usepackage{amssymb}
\usepackage{amsfonts}
\usepackage{mathrsfs}
\usepackage{xspace}
\usepackage{bm}
\usepackage{fancyref}
\usepackage{textcomp}

\usepackage{multirow}
\usepackage{stmaryrd}

\newenvironment{textbmatrix}{	\setlength{\arraycolsep}{2.5pt}%
								\big[\begin{matrix}}{\end{matrix}\big]%
								\raisebox{0.08ex}{\vphantom{M}}}

\def\be{\begin{equation}}
\def\ee{\end{equation}}
\def\een{\nonumber \end{equation}}
\def\mat{\begin{bmatrix}}
\def\emat{\end{bmatrix}}
\def\btm{\begin{textbmatrix}}
\def\etm{\end{textbmatrix}}

\def\ba#1\ea{\begin{align}#1\end{align}}
\def\bas#1\eas{\begin{align*}#1\end{align*}}
\def\bs#1\es{\begin{split}#1\end{split}} 
\def\bg#1\eg{\begin{gather}#1\end{gather}}
\def\bml#1\eml{\begin{multline}#1\end{multline}}
\def\bi#1\ei{\begin{itemize}#1\end{itemize}}

\newcommand{\subtext}[1]{\text{\fontfamily{cmr}\fontshape{n}\fontseries{m}\selectfont{}#1}}
\newcommand{\lefto}{\mathopen{}\left}

\DeclareMathOperator{\Exop}{\opE}			
\DeclareMathOperator{\Varop}{\mathrm{Var}}
\DeclareMathOperator{\Covop}{\mathrm{Cov}}
\DeclareMathOperator{\rot}{\nabla\times}		

\newcommand{\Ex}[2]{\ensuremath{\Exop_{#1}\lefto[#2\right]}} 	
\newcommand{\abs}[1]{\lefto\lvert#1\right\rvert}		

\newcommand{\vecnorm}[1]{\lefto\lVert#1\right\rVert}		

\safemath{\dirac}{\delta}					
\safemath{\krond}{\dirac}					

\safemath{\upto}{\uparrow}
\safemath{\downto}{\downarrow}
\safemath{\iu}{j}							
\safemath{\ev}{\lambda}						
\safemath{\hilseqspace}{l^{2}}				
\newcommand{\banachfunspace}[1]{\setL^{#1}}	
\safemath{\hilfunspace}{\banachfunspace{2}}	

\safemath{\SNR}{\textsf{SNR}} 				
\safemath{\PAR}{\textsf{PAR}} 				
\safemath{\No}{N_0}							
\safemath{\Es}{E_s}							
\safemath{\Eb}{E_b}							
\safemath{\EbNo}{\frac{\Eb}{\No}}
\safemath{\EsNo}{\frac{\Es}{\No}}

\DeclareMathOperator{\CHop}{\ensuremath{\opH}} 
\safemath{\tvir}{\rndh_{\CHop}}				
\safemath{\tvtf}{\rndl_{\CHop}}				
\safemath{\spf}{\rnds_{\CHop}}				
\safemath{\bff}{H_{\CHop}}					

\safemath{\ircf}{r_{h}}						
\safemath{\tftvcf}{r_{s}}					
\safemath{\tfcf}{r_{l}}						
\safemath{\bfcf}{r_{H}}						

\safemath{\tcorr}{c_h}						
\safemath{\scf}{c_{s}}						
\safemath{\tfcorr}{c_{l}}					
\safemath{\fcorr}{c_{H}}						

\safemath{\mi}{I}							
\safemath{\capacity}{C}						

\safemath{\normal}{\mathcal{N}}			
\safemath{\jpg}{\mathcal{CN}}			
\safemath{\mchain}{\leftrightarrow}		
\newcommand{\AIC}{\ensuremath{\text{AIC}}} 

\safemath{\dB}{\,\mathrm{dB}}
\safemath{\dBm}{\,\mathrm{dBm}}
\safemath{\Hz}{\,\mathrm{Hz}}
\safemath{\kHz}{\,\mathrm{kHz}}
\safemath{\MHz}{\,\mathrm{MHz}}
\safemath{\GHz}{\,\mathrm{GHz}}
\safemath{\s}{\,\mathrm{s}}
\safemath{\ms}{\,\mathrm{ms}}
\safemath{\mus}{\,\mathrm{\text{\textmu}s}}
\safemath{\ns}{\,\mathrm{ns}}
\safemath{\ps}{\,\mathrm{ps}}
\safemath{\meter}{\,\mathrm{m}}
\safemath{\mm}{\,\mathrm{mm}}
\safemath{\cm}{\,\mathrm{cm}}
\safemath{\m}{\,\mathrm{m}}
\safemath{\W}{\,\mathrm{W}}
\safemath{\mW}{\, \mathrm{mW}}
\safemath{\J}{\,\mathrm{J}}
\safemath{\K}{\,\mathrm{K}}
\safemath{\bit}{\,\mathrm{bit}}
\safemath{\nat}{\,\mathrm{nat}}

\safemath{\define}{\triangleq}			

\safemath{\equivalent}{\sim}
\safemath{\distas}{\sim}					
\safemath{\sdiff}{\Delta}				

\safemath{\reals}{\mathbb{R}}
\safemath{\positivereals}{\reals_{+}}
\safemath{\integers}{\mathbb{Z}}
\safemath{\posint}{\integers_{+}}
\safemath{\naturals}{\mathbb{N}}
\safemath{\posnaturals}{\naturals_{+}}
\safemath{\complexset}{\mathbb{C}}
\safemath{\rationals}{\mathbb{Q}}

\newcommand*{\fancyrefapplabelprefix}{app}		
\newcommand*{\fancyrefthmlabelprefix}{thm}		
\newcommand*{\fancyreflemlabelprefix}{lem}		
\newcommand*{\fancyrefcorlabelprefix}{cor}		
\newcommand*{\fancyrefdeflabelprefix}{def}		
\newcommand*{\fancyrefproplabelprefix}{prop}	
\newcommand*{\fancyrefobslabelprefix}{obs}		
\newcommand*{\fancyrefalglabelprefix}{alg}		
\newcommand*{\fancyrefasmlabelprefix}{asm}	    
\newcommand*{\fancyreftbllabelprefix}{tab}	 

\frefformat{vario}{\fancyrefseclabelprefix}{Section~#1}
\frefformat{vario}{\fancyrefthmlabelprefix}{Theorem~#1}
\frefformat{vario}{\fancyreflemlabelprefix}{Lemma~#1}
\frefformat{vario}{\fancyrefcorlabelprefix}{Corollary~#1}
\frefformat{vario}{\fancyrefdeflabelprefix}{Definition~#1}
\frefformat{vario}{\fancyrefobslabelprefix}{Observation~#1}
\frefformat{vario}{\fancyrefasmlabelprefix}{Assumption~#1}
\frefformat{vario}{\fancyreffiglabelprefix}{Fig.~#1}
\frefformat{vario}{\fancyrefapplabelprefix}{Appendix~#1} 
\frefformat{vario}{\fancyrefproplabelprefix}{Proposition~#1}
\frefformat{vario}{\fancyrefalglabelprefix}{Algorithm~#1}
\frefformat{vario}{\fancyreftbllabelprefix}{Table~#1}
\frefformat{vario}{\fancyrefeqlabelprefix}{(#1)}

\newcommand{\splitatcommas}[1]{%
  \begingroup
  \ifnum\mathcode`,="8000
  \else
    \begingroup\lccode`~=`, \lowercase{\endgroup
      \edef~{\mathchar\the\mathcode`, \penalty0 \noexpand\hspace{0pt plus 1em}}%
    }\mathcode`,="8000
  \fi
  #1%
  \endgroup
}

\begin{document}

\title{Distortion-Aware Linear Precoding for Millimeter-Wave Multiuser MISO~Downlink}
\author{\IEEEauthorblockN{Sina Rezaei Aghdam$^\text{1}$, Sven Jacobsson$^\text{1,2}$, and Thomas Eriksson$^\text{1}$} \\[-0.3cm]
\thanks{The work of Sina Rezaei Aghdam and Thomas Eriksson was performed within the strategic innovation program ``Smarter Electronics Systems'', a common venture by VINNOVA, Formas, and Energimyndigheten.
The work of Sven Jacobsson was supported in part by the Swedish Foundation for Strategic Research under grant ID14-0022, and by VINNOVA within the competence center ChaseOn.}
\IEEEauthorblockA{
\small $^\text{1}$\textit{Chalmers University of Technology, Gothenburg, Sweden};  
\,$^\text{2}$\textit{Ericsson Research, Gothenburg, Sweden}
}
}



\maketitle

\begin{abstract}
In this work, we propose an iterative scheme for computing a linear precoder that takes into account the impact of hardware impairments in the multiuser multiple-input single-output downlink. We particularly focus on the case when the transmitter is equipped with nonlinear power amplifiers. Using Bussgang's theorem, we formulate a lower bound on the achievable sum rate in the presence of hardware impairments, and maximize it using projected gradient ascent. We provide numerical examples that demonstrate the efficacy of the proposed distortion-aware scheme for precoding over a millimeter-wave~channel.
\end{abstract}


\section{Introduction}

Wireless communication over millimeter-wave (mmWave) frequency bands combined with large-scale multi-antenna transmission techniques promises significant improvements in spectral efficiency compared to today's state-of-the-art communication systems~\cite{swindlehurst14a}. These technologies are believed to be key enablers for future communication systems, including fifth generation~(5G) cellular networks~\cite{boccardi14a}.
Recently, a large body of research has been conducted on studying the potentials of mmWave multi-antenna transmission schemes (see, e.g., \cite{busari18a} for a survey). However, the vast majority of these works relies on the assumption of ideal transceiver hardware, which is not a valid assumption in realistic systems.

In practice, the performance of multi-antenna systems is limited by different transceiver hardware impairments such as amplifier nonlinearities, phase noise, in-phase/quadrature (I/Q) imbalance, and quantization noise. 
Modeling of these impairments and evaluating the performance loss imposed by them has been a topic of much recent interest. 
Existing studies in this area can be categorized into two groups. 
The first group of works is focused on the impact of a single (or pre-dominant) hardware impairment. For example, the impact of power amplifier (PA) nonlinearities on the performance of multi-antenna systems has been investigated in, e.g.,~\cite{qi12a,blandino17a,mollen18b}. The work in
~\cite{moghadam18a} characterizes the performance of mmWave multi-antenna systems (in terms of spectral and energy efficiency) in the presence of PA nonlinearities and crosstalk.
The impact of other hardware impairments such as phase noise, I/Q imbalance, and quantization has been investigated in, e.g.,~\cite{mezghani12b, khanzadi15a, kolomvakis16a, jacobsson17d, de-candido18a}.
The second group of works is concerned with evaluating the aggregate impact of several hardware impairments~(see, e.g., \cite{studer10b, zhang15d, bjornson14a}).
In these works, the distortion caused by nonideal hardware is modeled as an additive Gaussian noise that is uncorrelated over the antenna array. 
This is, however, not a realistic assumption as the distortion caused by nonlinearities is, in general, correlated over the antenna array~\cite{larsson18a,bjornson19a}. 
More recently, an aggregate hardware-impairment model for the distortion caused by nonlinear amplifiers, phase noise, and quantization, which captures the inherent correlation within the distortion, was provided in~\cite{jacobsson18d}.


In this work, we propose an iterative scheme, which we refer to as distortion-aware beamforming (DAB), for finding a \emph{linear} precoder that takes into account the use of nonlinear PAs at the transmitter.
Specifically, we consider a downlink mmWave multiuser multiple-input single-output (MISO) system and formulate a non-convex optimization problem to find the linear precoder that maximizes a lower bound on the sum rate, which we solve approximately using gradient ascent.
We demonstrate the efficacy of the proposed DAB precoder by means of numerical simulation. 
Specifically, we show that, by taking nonlinear distortion into account, the DAB precoder outperforms conventional maximal-ratio transmission (MRT) and zero-forcing~(ZF) precoding.
%

The rest of the paper is organized as follows. In Section~\ref{sec:SM}, we introduce the system model and formulate the sum-rate optimization problem in the presence of nonlinearities at the transmitter. In Section \ref{sec:GP}, we present the iterative scheme used to compute the DAB precoding matrix. In Section \ref{sec:Num}, we examine the efficacy of the proposed precoding scheme via numerical examples. We conclude the paper in Section~\ref{sec:Conc}.

Lowercase and uppercase boldface letters denote vectors and matrices, respectively. 
The superscripts $(\cdot)^*$, $(\cdot)^T$, and $(\cdot)^H$ denote complex conjugate, transpose, and Hermitian transpose, respectively.
We use $\mathbb{E}[\cdot]$ to denote expectation. We use $\vecnorm{\veca}$ to denote the $\ell_2$-norm of $\veca$.
The $M \times M$ identity matrix is denoted by $\matI_M$ and the $M \times M$ all-zeros matrix is denoted by $\matzero_{M \times M}$. 
We use $\matA \odot \matB$ to denote the Hadamard (entry-wise) product of two equally-sized matrices $\matA$ and $\matB$.
%
%
Moreover, $\text{diag}(\matA)$ is the main diagonal of a square matrix $\matA$. 
The distribution of a circularly-symmetric complex Gaussian random vector with covariance matrix $\matC \in \opC^{M \times M}$ is denoted by $\mathcal{CN}(\boldsymbol{0}_{M \times M}, \matC)$.
Finally, we use $\mathds{1}_{\setA}(a)$ to denote the indicator function, which is defined as~$\mathds{1}_{\setA}(a) = 1$ for $a \in \setA$ and $\mathds{1}_{\setA}(a) = 0$ for~$a \notin \setA$.

\section{System Model and Problem Formulation} \label{sec:SM}

\begin{figure}[t]
	\centering
	\includegraphics[width=.95\columnwidth]{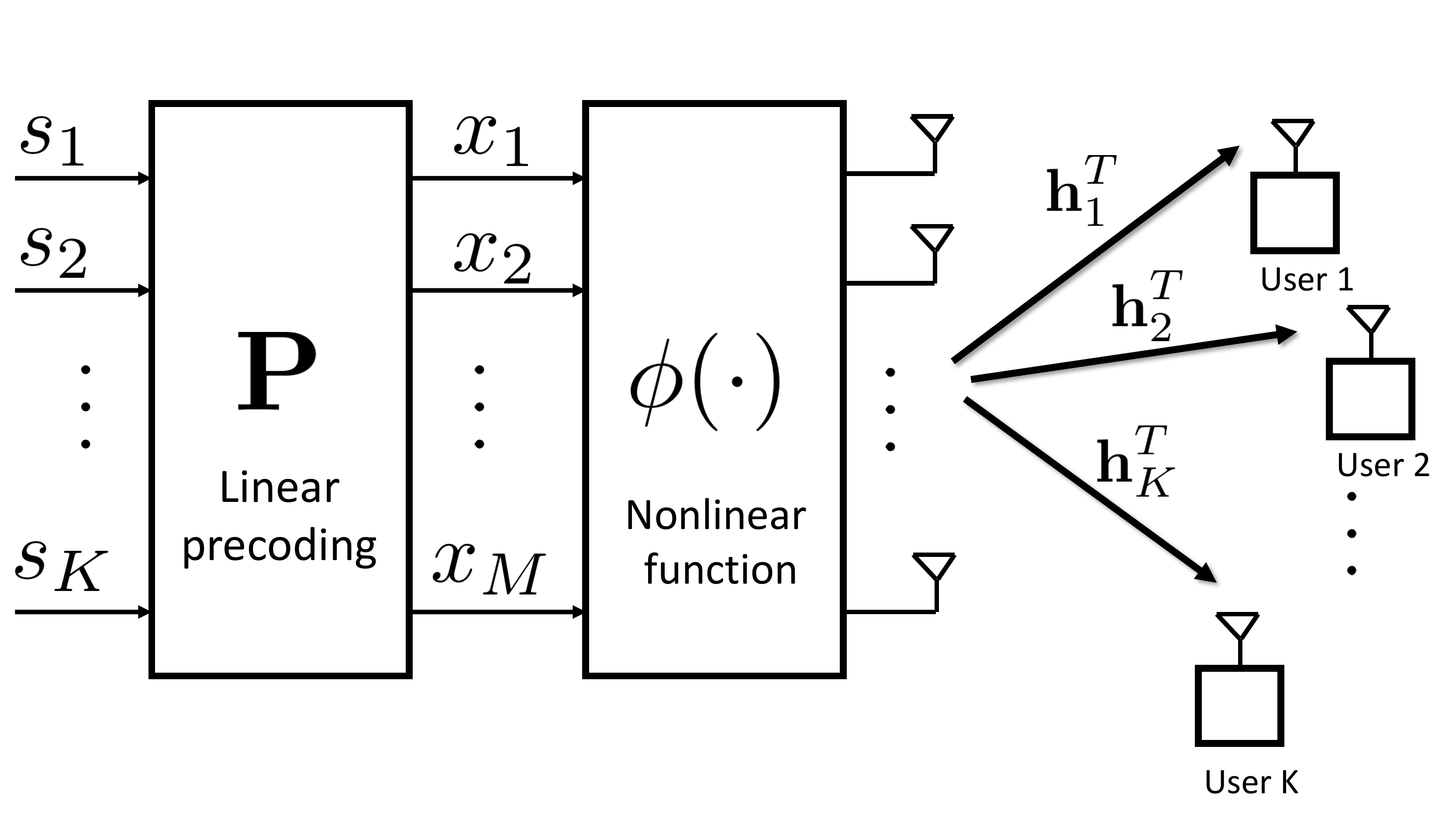}
	\caption{Multiuser MISO downlink with linear precoding and hardware impairments at the transmitter.}
	\label{system_model}
\end{figure}

We consider the nonlinearly distorted multiuser MISO system depicted in Fig.~\ref{system_model}. Here, an $M$-antenna transmitter serves $K$ single-antenna users in the same time-frequency resource. The received signal at the $k$th user is given~by
\begin{equation} \label{Received}
y_k = \vech_k^T \phi(\vecx) + w_k,
\end{equation}
for $k = 1,\dots,K$.
Here, $\vech_k \in \opC^{M}$ is the channel between the transmitter and the $k$th user (which we assume is constant for the duration of each codeword), $\vecx = \left[x_1, \dots, x_{M} \right]^T\in \opC^{M}$ is the precoded vector, and $w_k  \sim \mathcal{CN}(0, N_0)$ is the additive white Gaussian noise (AWGN). 
We use the nonlinear function $\phi(\cdot): \opC \rightarrow \opC$, which is applied entry-wise on a vector, to model the nonlinear PAs at the transmitter.

We consider linear precoding such that $\vecx = \matP \vecs$, where $\matP = [\vecp_1, \dots, \vecp_K] \in \mathbb{C}^{M \times K}$ is the precoding matrix and $\vecs = [s_1, \dots, s_K]^T \sim \mathcal{CN}(\matzero_{K \times K}, \matI_{K})$ are the transmitted symbols.

\subsection{Modeling of Transmitter Hardware Impairments}

In order to analyze the impact of the nonlinear distortion on the performance of the system, we shall, similarly to, e.g.,~\cite{mezghani12b, jacobsson18d, bjornson19a}, use Bussgang's theorem~\cite{bussgang52a}, which allows us to write the nonlinearly distorted signal $\phi(\vecx)$ as
\begin{equation}\label{Bussgang}
\phi(\vecx) = \matB \vecx + \vece, 
\end{equation}
where the distortion term $\vece \in \opC^M$ is uncorrelated with $\vecx$, i.e., $\mathbb{E}[\vecx\vece^H] = \matzero_{M \times M}$. Furthermore, $\matB \in \opC^{M \times M}$ is a diagonal matrix whose entries along the diagonal are given by $[\matB]_{m,m} = {\Ex{}{\phi(x_m) x_m^*}} / {\Ex{}{|x_m|^2}}$ for $m = 1,\dots,M$. 
In this work, for simplicity, we shall use a third-order polynomial model for the nonlinear PAs at the transmitter.~Specifically,
\begin{equation} \label{eq:third_order}
\phi(x) = \beta_1 x + \beta_3 x |x|^2,
\end{equation} 
where $ \beta_1 \in \mathbb{C}$ and $\beta_3 \in \mathbb{C}$ are the model parameters, which we assume are known to the transmitter. A possible way of acquiring such knowledge is by performing over-the-air measurements using one or few observation receivers at the transmitter~(see, e.g., \cite{hausmair18a, hesami19a}). 
For the third-order polynomial model in~\eqref{eq:third_order} and for $\vecx = \matP\vecs$, it holds that the gain matrix $\matB$ in~\eqref{Bussgang} depends on the precoding matrix $\matP$~as 
\begin{equation}
\matB(\matP) = \beta_1\matI_M + 2\beta_3 \text{diag}(\matP\matP^H).
\end{equation}

\subsection{An Achievable Sum Rate}

By inserting (\ref{Bussgang}) into (\ref{Received}), the received signal at the $k$th user can be written as
\begin{IEEEeqnarray}{rCl} 
y_k 
&=&  \vech_k^T \matB(\matP)  \vecp_k s_k + \sum_{r \neq k} \vech^T_k \matB(\matP)\vecp_r s_r + \vech_k^T \vece + w_k, \IEEEeqnarraynumspace \label{eq:received_at_user}
\end{IEEEeqnarray}
where $s_k$ is the desired symbol at the $k$th user.
It should be noted that the effective noise term $\sum_{r \neq k} \vech^T_k \matB(\matP) \vecp_r s_r + \vech_k^T \vece + w_k$ in~\eqref{eq:received_at_user} is, in general, non-Gaussian distributed due to the nonlinearity at the transmitter. 
Since Gaussian noise is the worst-case additive noise (in terms of mutual information) for Gaussian inputs under a covariance constraint~\cite{diggavi01a}, an achievable sum rate can be formulated~as
\begin{equation} \label{sum_rate}
R_{\text{sum}}(\matP) = \sum_{k = 1}^{K} \log_2 \left(1 + \text{SINDR}_k(\matP)\right),
\end{equation}
where $\text{SINDR}_k(\matP)$ is the signal-to-interference-noise and distortion ratio (SINDR) at the $k$th user, which is given by
\begin{IEEEeqnarray}{rCl}  \label{SINR}
\text{SINDR}_k(\matP) = \frac{|\vech^T_k \matB(\matP) \vecp_k|^2}{\sum\limits_{r \neq k} |\vech^T_k \matB(\matP) \vecp_r|^2 + \vech^T_k \matC_\vece(\matP) \vech_k^{*} + N_0}. \IEEEeqnarraynumspace
\end{IEEEeqnarray}
Here, $\matC_{\vece}(\matP) \in \opC^{M \times M}$ is the covariance of the distortion~$\vece$, which is given by~(see, e.g.,~\cite[Eq.~(24)]{bjornson19a})
\begin{equation}
\matC_\vece(\matP) = 2 \abs{\beta_3}^2 \lefto(\matP \matP^H  \odot \matP^*\matP^T \odot \matP\matP^H \right).
\end{equation}

\subsection{The Optimization Problem}

Clearly, the choice of precoding matrix $\matP$ has an impact on the sum rate in \eqref{sum_rate}. 
Under the assumption of perfect channel state information (CSI) at the transmitter, our objective is to find the precoding matrix $\matP$ that maximizes the sum rate in~\eqref{sum_rate} under an equality constraint $\mathbb{E}[\vecnorm{\phi(\matP\vecs)}^2] = P_\text{tot}$ on the average transmit power. This optimization problem can be formulated as~follows:
\begin{align} \label{MaxRsum}
\begin{array}{lll}
\underset{\matP\in \opC^{M \times K}}{\text{maximize}} & R_{\text{sum}}(\matP)\\
\text{subject to} & \Ex{}{\vecnorm{\phi(\matP\vecs)}^2} = P_{\text{tot}}.
\end{array}
\end{align}
Note that~\eqref{MaxRsum} is a non-convex optimization problem since $R_{\text{sum}}(\matP)$ is a non-convex function of $\matP$. Next, we shall solve this problem approximately using the iterative algorithm described in~Section~\ref{sec:GP}.

\begin{algorithm}[b!]
	\begin{spacing}{1.2}
	\caption{Algorithm for computing the distortion-aware beamforming (DAB) precoding matrix.}
	\begin{algorithmic}[1]
		\renewcommand{\algorithmicrequire}{\textbf{Inputs:}}
		\renewcommand{\algorithmicensure}{\textbf{Output:}}
		\REQUIRE $\vech_1, \dots, \vech_K$, $\beta_1$, $\beta_3$, $P_\text{tot}$, and $N_0$ 
		\ENSURE  $\matP_\text{DAB}$
		\\ \textit{Initialization}: $\mu^{(0)}$ and $\matP^{(0)}$
		\STATE $R_\text{sum}^{(0)} \leftarrow R_{\text{sum}}(\matP^{(0)})$
		\FOR{$i = 1,\dots,I$}
		\STATE $ \widetilde\matP \leftarrow \lefto[ \matP^{(i-1)} \!+\! \mu^{(i-1)} \nabla_\matP R_{\text{sum}}\big(\matP^{(i-1)}\big) \right]_{\opE{\vecnorm{\phi\lefto(\matP\vecs\right)}^2} = P_{\text{tot}}}^{+}$
		\STATE $\widetilde{R}_\text{sum} \leftarrow R_{\text{sum}}(\widetilde\matP)$
		\IF {$\widetilde{R}_\text{sum} > R_\text{sum}^{(i-1)}$}
		\STATE $\matP^{(i)} \leftarrow \widetilde\matP$, $R_\text{sum}^{(i)} \leftarrow \widetilde{R}_\text{sum}$, and $\mu^{(i)} \leftarrow \mu^{(0)}$.
		\ELSE
		\STATE $\matP^{(i)} \leftarrow \matP^{(i-1)}$, $R_\text{sum}^{(i)} \leftarrow R_\text{sum}^{(i-1)}$, and $\mu^{(i)} \leftarrow \frac{1}{2}\mu^{(i-1)}$
		\ENDIF
		\ENDFOR
		\STATE $\matP_\text{DAB} \leftarrow \matP^{(I)}$
	\end{algorithmic} 
	\end{spacing}
\end{algorithm}

\section{Distortion-Aware Linear Precoding} \label{sec:GP}

In what follows, we solve the constrained non-convex optimization problem \eqref{MaxRsum} approximately using an iterative scheme based on gradient ascent followed by a projection step to ensure the feasibility of the solution. We shall refer to the output of the iterative scheme as the DAB precoding matrix.
Specifically, our iterative solution updates the precoding matrix by taking steps along the steepest ascent direction of the objective function $R_{\text{sum}}(\matP)$ followed by normalization of the resulting precoding matrix as follows:
\begin{IEEEeqnarray}{rCl} \label{eq:update_precoding_matrix}
\widetilde\matP &=& \lefto[ \matP^{(i-1)} + \mu^{(i-1)} \nabla_\matP R_{\text{sum}}\big(\matP^{(i-1)}\big) \right]_{\opE{\vecnorm{\phi\lefto(\matP\vecs\right)}^2} = P_{\text{tot}}}^{+}\!. \IEEEeqnarraynumspace
\end{IEEEeqnarray}
Here, $i = 1,\dots, I$ is the iteration index, $I$ is the maximum number of iterations, $\mu^{(i)}$ is the step size of the $i$th iteration, and $[\cdot]_{\opE{\vecnorm{\phi\lefto(\matP\vecs\right)}^2} = P_{\text{tot}}}^{+}$ denotes normalization of the updated precoding matrix such that the power constraint in (\ref{MaxRsum}) is satisfied.
If $R_\text{sum}(\widetilde\matP) > R_\text{sum}(\matP^{(i-1)})$, we update the precoding matrix to $\matP^{(i)} = \widetilde\matP$ and reset the step size $\mu^{(i)} = \mu^{(0)}$. Otherwise, we do not update the precoding matrix, i.e., $\matP^{(i)} = \matP^{(i-1)}$, and decrease the step size $\mu^{(i)} = \frac{1}{2}\mu^{(i-1)}$. Finally, we choose $\matP_\text{DAB} = \matP^{(I)}$ as the DAB precoding matrix.
In Algorithm~1, we summarize the steps required for computing the DAB precoding matrix using the projected gradient~ascent approach.

Next, we shall provide a closed-form expression for the gradient $\nabla_{\mathbf{P}} R_\text{sum}(\mathbf{P}) \in \opC^{M \times K}$, which is required to evaluate the update step~\eqref{eq:update_precoding_matrix}. To this end, let
\begin{IEEEeqnarray}{rCl} \label{eq:numerator}
	n_k(\mathbf{P}) &=& 	|\mathbf{h}^T_k \mathbf{B}(\mathbf{P}) \mathbf{p}_k|^2,
\end{IEEEeqnarray}
denote the numerator of the SINDR in~\eqref{SINR}. Furthermore, let 
\begin{IEEEeqnarray}{rCl} \label{eq:denom}
	d_k(\mathbf{P}) &=& 	 d^\text{\,mui}_k(\matP)  + d^\text{\,dist}_k(\matP)  + N_0,
\end{IEEEeqnarray}
denote the denominator of the SINDR in~\eqref{SINR}, where $d^\text{\,mui}_k(\matP) = \sum_{r \neq k} |\vech^T_k \matB(\matP) \vecp_r|^2$ and $d^\text{\,dist}_k(\matP) = \vech^T_k \matC_{\vece}(\matP) \vech_k^{*}$ is the part of the denominator corresponding to multiuser interference and nonlinear distortion, respectively.
With these definitions, the gradient $\nabla_{\matP} R_\text{sum}(\matP)$ can be written as
\begin{IEEEeqnarray}{rCl}
\nabla_\matP{R}_\text{sum}(\matP)
&=& \sum_{k=1}^K \frac{2 \log_2(e)}{{d_k^2(\matP)(1 + n_k(\matP)/d_k(\matP))}} \nonumber\\
&& \!\times \lefto( d_k(\matP)\frac{\partial n_k(\matP)}{\partial\matP^*}- n_k(\matP) \frac{\partial d_k(\matP)}{\partial\matP^*}\right), \label{eq:gradient_of_sum_rate} \IEEEeqnarraynumspace
\end{IEEEeqnarray}
where $\partial n_k(\matP) / \partial\matP^* = \splitatcommas{[\partial{n_k}(\matP)/\partial\vecp_1^*, \dots, \partial{n_k}(\matP)/\partial\vecp_K^*]}$ and $\partial d_k(\matP) / \partial\matP^* = \splitatcommas{[\partial{d_k}(\matP)/\partial\vecp_1^*, \dots, \partial{d_k}(\matP)/\partial\vecp_K^*]}$. 
Hence, to compute the gradient $\nabla_{\mathbf{P}} R_\text{sum}(\mathbf{P})$, we need to compute the derivatives of $n_k(\matP)$ and $d_k(\matP)$ for $k = 1,\dots,K$.
Starting with the numerator, it can be shown~that the derivative with respect to $\vecp_{k'}^*$ can be written~as
\begin{IEEEeqnarray}{rCl} \label{eq:deriv_numerator}
\frac{\partial{n_k}(\matP)}{\partial\vecp_{k'}^*} 
&=& \lefto(\mathbf{\Gamma}_{k}(\matP) \mathds{1}_{\{{k'} = k\}}(k') + \mathbf{\Upsilon}_{k,k'}(\matP)\right)\vecp_{k'}, 
\end{IEEEeqnarray}
for $k' = 1,\dots,K$. Here, we have defined $\mathbf{\Gamma}_{k}(\matP) \in \opC^{M \times M}$~as
\begin{IEEEeqnarray}{rCl}
\mathbf{\Gamma}_{k}(\matP) 
&=&  \abs{\beta_1}^2   \vech_k^*\vech_k^T  \nonumber\\
&& + 2 \lefto(\beta_1^*\beta_3  \vech_k^*\vech_k^T \text{diag}(\matP\matP^H) \right.\nonumber\\
&& + \lefto. \beta_1\beta_3^* \text{diag}(\matP\matP^H) \vech_k^*\vech_k^T\right) \IEEEeqnarraynumspace\nonumber\\  
&& + 4\abs{\beta_3}^2 \text{diag}(\matP\matP^H)\vech_k^*\vech_k^T\text{diag}(\matP\matP^H).
\end{IEEEeqnarray}
Furthermore, we have defined $\mathbf{\Upsilon}_{k,k'}(\matP) \in \opC^{M \times M}$ as
\begin{IEEEeqnarray}{rCl}
\mathbf{\Upsilon}_{k,k'}(\matP)
&=& 2 \lefto( \beta_1^*\beta_3 \text{diag}\lefto( \vecp_{k'}\vecp_{k'}^H \vech_k^*\vech_k^T \right) \right.\nonumber\\
&& \lefto. +  \beta_1\beta_3^* \text{diag}\lefto( \vech_k^*\vech_k^T\vecp_{k'}\vecp_{k'}^H \right)\right) \nonumber\\
&& +  4\abs{\beta_3}^2 \lefto( \text{diag}\lefto( \vech_k^*\vech_k^T \text{diag}(\matP\matP^H) \vecp_{k'}\vecp_{k'}^H \right) \right. \IEEEeqnarraynumspace \nonumber\\
&& \lefto. +  \text{diag}\lefto( \vecp_{k'}\vecp_{k'}^H \text{diag}(\matP\matP^H) \vech_k^*\vech_k^T  \right) \right).
\end{IEEEeqnarray}
The derivative with respect to $\vecp_{k'}^*$ of the denominator can be written as
\begin{IEEEeqnarray}{rCl} \label{eq:deriv_denom}
	\frac{\partial{d_k}(\matP)}{\partial\vecp_{k'}^*} &=& 	\frac{\partial{d^\text{\,mui}_k}(\matP)}{\partial\vecp_{k'}^*} + \frac{\partial{d^\text{\,dist}_k}(\matP)}{\partial\vecp_{k'}^*},
\end{IEEEeqnarray}
where the derivative of the multiuser-interference term in the denominator is given~by
\begin{IEEEeqnarray}{rCl} \label{eq:deriv_denom_mui}
\frac{\partial{d^\text{\,mui}_k}(\matP)}{\partial\vecp_{k'}^*}
&=& \lefto(\mathbf{\Gamma}_{k}(\matP) \mathds{1}_{\{{k'} \neq k\}}(k') + \sum_{r \neq k} \mathbf{\Upsilon}_{k,r}(\matP)\right)\!\vecp_{k'}, \IEEEeqnarraynumspace
\end{IEEEeqnarray}
for $k' = 1, \dots, K$. Furthermore, the $m$th entry of the derivative of the nonlinear-distortion term in the denominator is given by
\begin{IEEEeqnarray}{rCl} \label{eq:deriv_denom_dist}
\frac{\partial{d^\text{\,dist}_k}(\matP)}{\partial{p^*_{m,k'}}}
&=& 2 \abs{\beta_3}^2 \! \lefto( 2 h^*_{k,m} \!\! \sum_{m'=1}^M \! h_{k,m'} p_{m',k'} \lefto[ \abs{\matP\matP^H}^2\right]_{m',m} \right. \quad  \nonumber\\
&& + \lefto. h_{k,m} \!\! \sum_{m'=1}^M \! h_{k,m'}^* p_{m',k'}\lefto[ \lefto(\matP\matP^H\right)^2\right]_{m,m'} \right), \IEEEeqnarraynumspace 
\end{IEEEeqnarray}
for $k' = 1, \dots, K$ and $m = 1, \dots, M$, where $h_{k,m} = [\vech_k]_m$ and $p_{m,k} = [\vecp_k]_m$.
Finally, by inserting \eqref{eq:numerator}, \eqref{eq:denom}, \eqref{eq:deriv_numerator}, \eqref{eq:deriv_denom}, \eqref{eq:deriv_denom_mui}, and \eqref{eq:deriv_denom_dist} into \eqref{eq:gradient_of_sum_rate}, we obtain a closed-form expression for the gradient $\nabla_{\mathbf{P}} 	R_\text{sum}(\mathbf{P})$.

Note that the objective function (i.e, the sum rate) in Algorithm 1 is nondecreasing from one iteration to the next and that, for a given SNR, it is bounded from above. Hence, convergence of this algorithm is guaranteed. 
In order to increase the likelihood of converging to the global maximum instead of local maximum, we repeat the algorithm with multiple initializations and pick the solution that achieves the highest sum~rate.
By including the MRT and ZF precoding matrices among the set of initializations, we can guarantee that the DAB precoder does not perform any worse than these conventional linear precoders.

\section{Numerical Examples} \label{sec:Num}

We verify the efficacy of the proposed DAB precoding scheme by means of numerical simulation. First, we adopt a geometric channel model with a few scatterers for which we evaluate the achievable sum rate as well as the convergence behavior of Algorithm 1. 
Second, we study the far-field radiation pattern of the transmitted signal, which provides some insight into the working principle of the proposed precoding~scheme.

In what follows, unless stated otherwise, we set $M = 16$ antennas, $\beta_1 = 0.98$, $\beta_3 = -0.04 - 0.01j$, and $P_\text{tot} = 43$\,dBm.
We set the number of iterations to $I =50$ and run Algorithm~1 for $50$ different initializations of $\matP^{(0)}$. Specifically, we initialize Algorithm 1 with the MRT and ZF precoding matrices along with $48$ random initializations (where the elements of $\matP^{(0)}$ are drawn from a Gaussian distribution).

\subsection{Geometric Channel Model} \label{sec:geo}

In order to capture the sparse scattering characteristics of mmWave channels in a non-line-of-sight (nLoS) environment, i.e., when there is no dominant path, we adopt a geometric channel model with $L$ scatterers as in, e.g.,\cite{alkhateeb14a, alkhateeb15a}, for~which
\begin{equation} \label{channel}
\vech_k = \sqrt{\frac{M}{L}} \sum_{\ell = 1}^{L} \alpha_{k,\ell} \veca(\psi_{k,\ell}),
\end{equation}
for $k = 1,\dots, K$. Here, $\alpha_{k,\ell} \sim \mathcal{CN}(0, \gamma^2)$ is the channel gain (including path loss) corresponding to the $\ell$th path, where $\gamma^2$ is the average path loss. 
Furthermore, $\psi_{k,\ell}$ is the angle of departure (AoD) for the $\ell$th path and $\veca(\psi_{k,\ell})$ is the corresponding array response vector. We assume that the transmit antennas are arranged in a uniform linear array (ULA) with $\lambda_c/2$ spacing (where $\lambda_c$ is the carrier wavelength) such that the $m$th entry of $\veca(\psi_{k,\ell})$~is
\begin{equation}
\lefto[\veca(\psi_{k,\ell})\right]_m = \frac{1}{\sqrt{M}}e^{-j  \pi(m-1) \cos(\psi_{k,\ell})},
\end{equation}
for $m = 1,\dots,M$. Throughout our simulations, we shall use the following definition of signal-to-noise ratio~(SNR):
\begin{equation}
\text{SNR} = \gamma^2  \frac{P_{\text{tot}} }{N_0}.
\end{equation}

\begin{figure}[!t]
	\centering	
	\subfloat[$K = 2$ users.]{\includegraphics[width = .91\columnwidth]{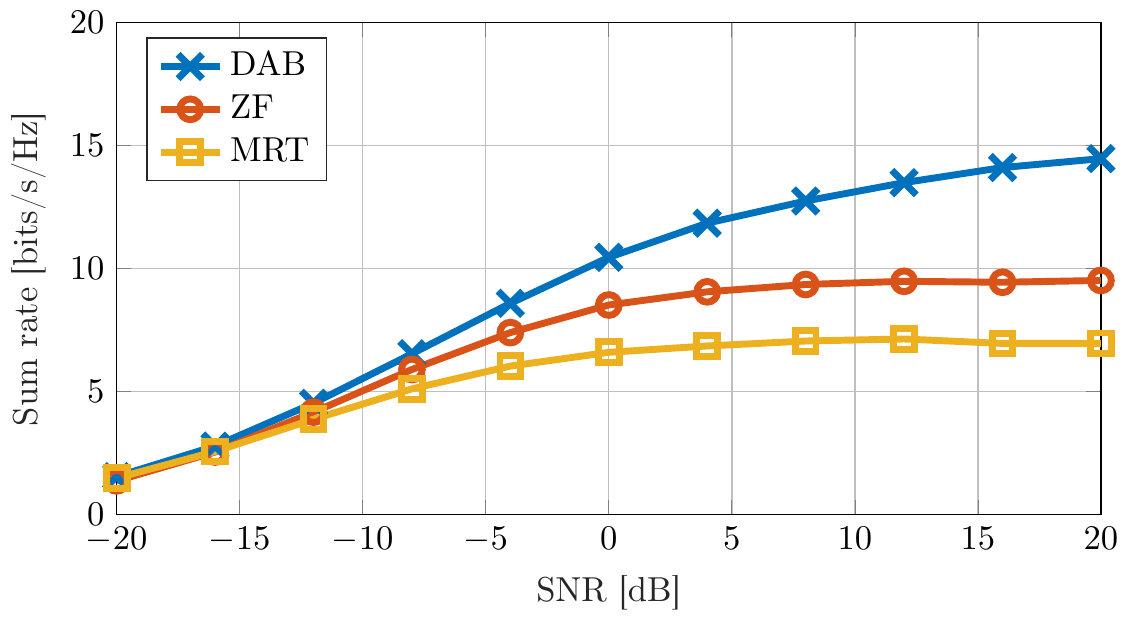}} \\
	\subfloat[$K = 4$ users.]{\includegraphics[width = .91\columnwidth]{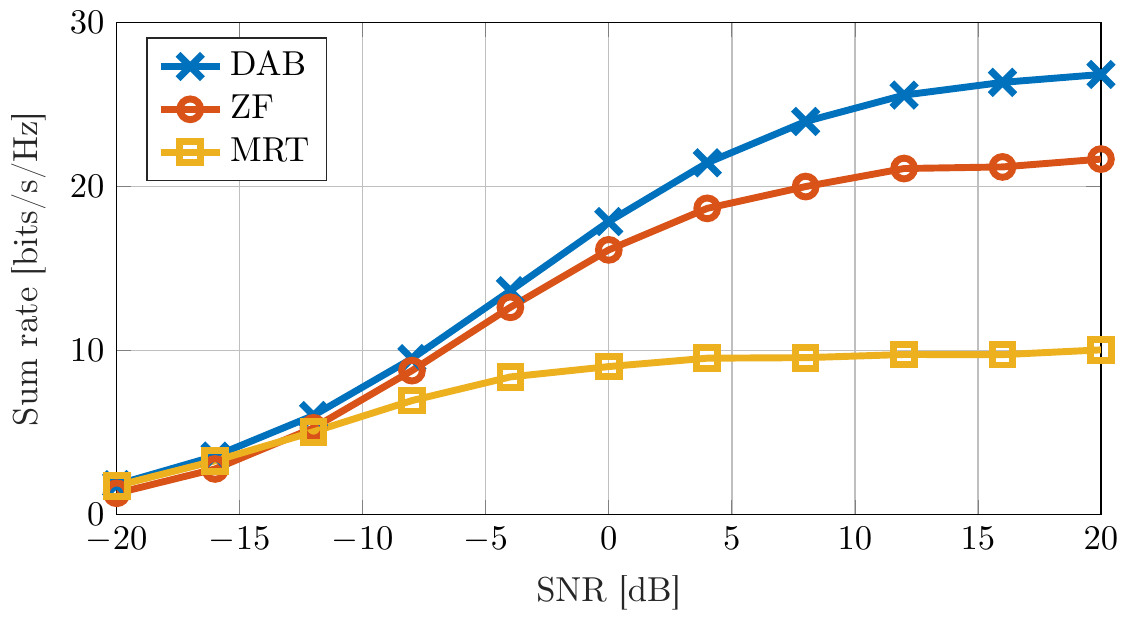}}
	\caption{Ergodic sum rate achievable with MRT, ZF, and DAB precoding; geometric channel model, $L=10$ paths, $M=16$ antennas, $\beta_1 = 0.98$, $\beta_3 = -0.04 - 0.01j$, and $P_\text{tot} = 43$\,dBm. The DAB precoder outperforms MRT and ZF~precoding.}
	\label{fig:sum_rate}
\end{figure}

\subsection{Performance Comparison} \label{sec:comparison}

In~\fref{fig:sum_rate}, we compare the sum rate, as a function of the SNR and the number of users, achieved by DAB precoding to the sum rate achieved by conventional MRT and ZF precoding.
We average the sum rate in~\eqref{sum_rate} over $10^3$ random channel realizations, generated using the model in \eqref{channel} under the assumption that the AoD $\psi_{k,\ell}$ is uniformly distributed over the interval $[0^\circ, 180^\circ]$.
By adopting the nLoS path-loss model presented in \cite[Table~I]{akdeniz14a} and by assuming that the system operates at carrier frequency $f_c = 28$\,GHz (such that $\lambda_c \approx 10.7$\,mm) with the average distance from the transmitter to the users being $20$\,m, we find that the average path loss is~$\gamma^2 = -110$\,dB.

We note from \fref{fig:sum_rate} that DAB precoding outperforms the conventional linear precoders that do not take into account the distortion introduced by the nonlinear PAs. The improved performance is particularly evident in the high-SNR regime for which the performance is limited by the nonlinear distortion. In the low-SNR regime, i.e., when the thermal noise dominates over the nonlinear distortion (and over the multiuser interference) the performance of the DAB precoder coincides with the performance of the MRT~precoder.
%

\subsection{Convergence of Algorithm 1}

The improved performance of the DAB precoder compared to conventional linear precoders is achieved at the cost of increased computational complexity. 
In particular, we obtain the DAB precoding matrix using the iterative scheme in Algorithm 1. Next, we study the convergence of Algorithm~1, which will provide some insight on the required number of iterations for achieving a certain performance.

\begin{figure}[!t]
	\centering	
	\includegraphics[width = .91\columnwidth]{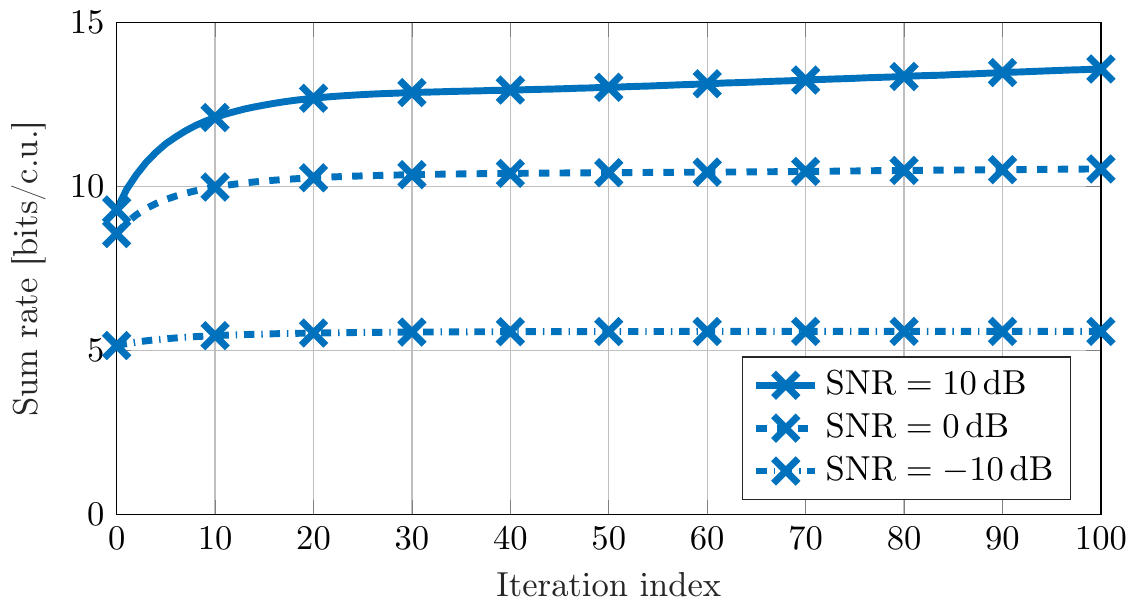}
	\caption{Average convergence behavior of Algorithm 1; geometric channel model, $L=10$ paths, $M=16$ antennas, $K = 2$ users, $\beta_1 = 0.98$, $\beta_3 = -0.04 - 0.01j$, and $P_\text{tot} = 43$\,dBm. The convergence is slower at high SNR compared to low SNR.}
	\label{fig:convergence}
\end{figure}

In \fref{fig:convergence}, we show the sum rate achieved by the DAB precoder as a function of the number of iterations. 
The rest of the simulation parameters are the same as in~\fref{sec:comparison}.
We note that the convergence of Algorithm~1 is faster at low SNR compared to high SNR.
Indeed, at low SNR, conventional precoding schemes are as good as DAB precoding (see \fref{fig:sum_rate}). 
Therefore, since the MRT and ZF precoding matrices are included in the initialization, fast convergence is expected. 
At high SNR, however, the structure of the optimal precoder is significantly different from the MRT and ZF precoding matrices, which results in slower convergence.

\begin{figure*}[t]
\centering
\subfloat[MRT; $K=1$ user and $\psi_{1} = 90^\circ$.]{\includegraphics[width =.44\textwidth]{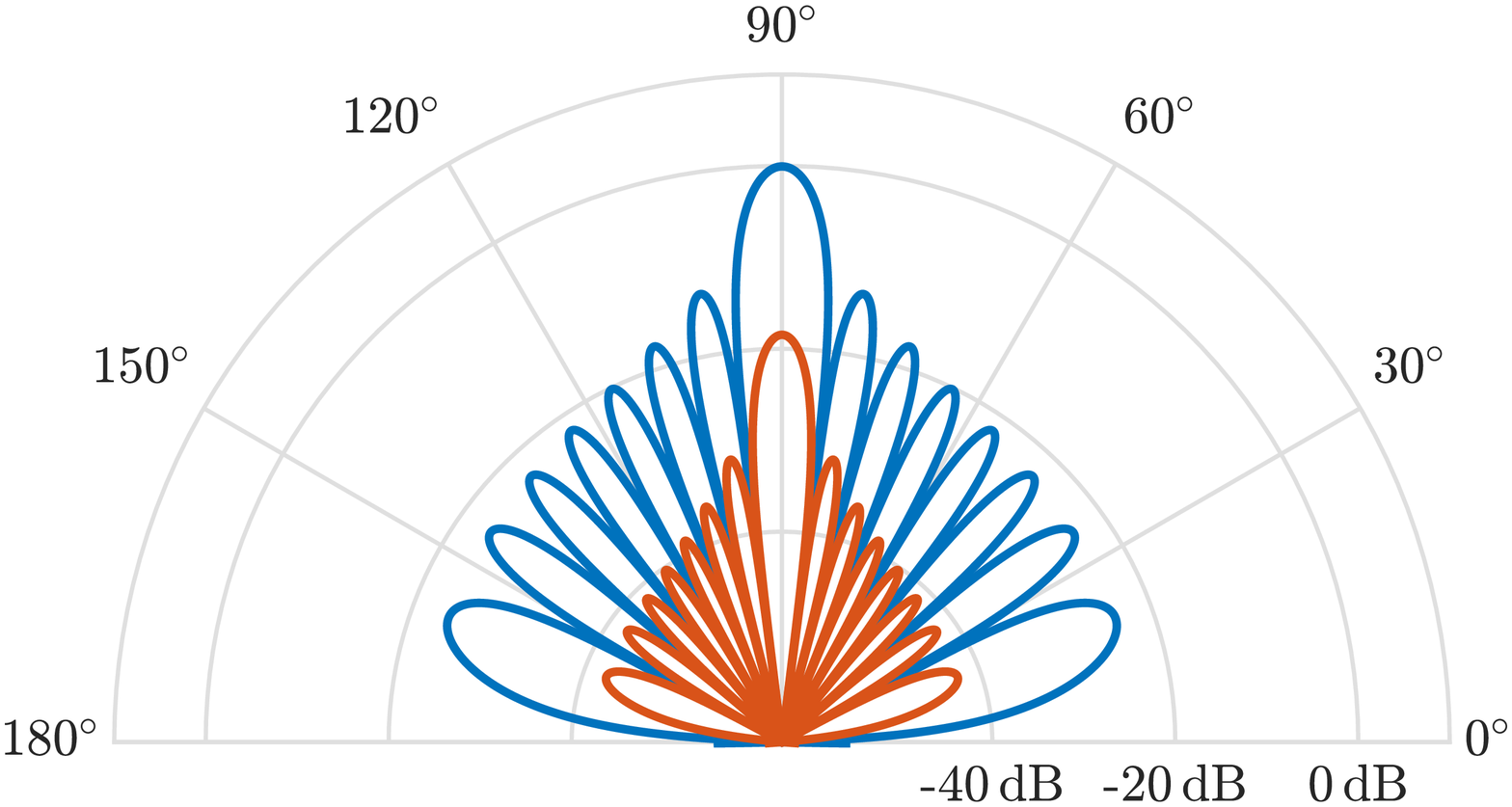}\label{fig:pattern_zf_1user}} \quad\quad
\subfloat[MRT; $K=2$ users, $\psi_{1} = 30^\circ$, and $\psi_{2} = 90^\circ$.]{\includegraphics[width =.44\textwidth]{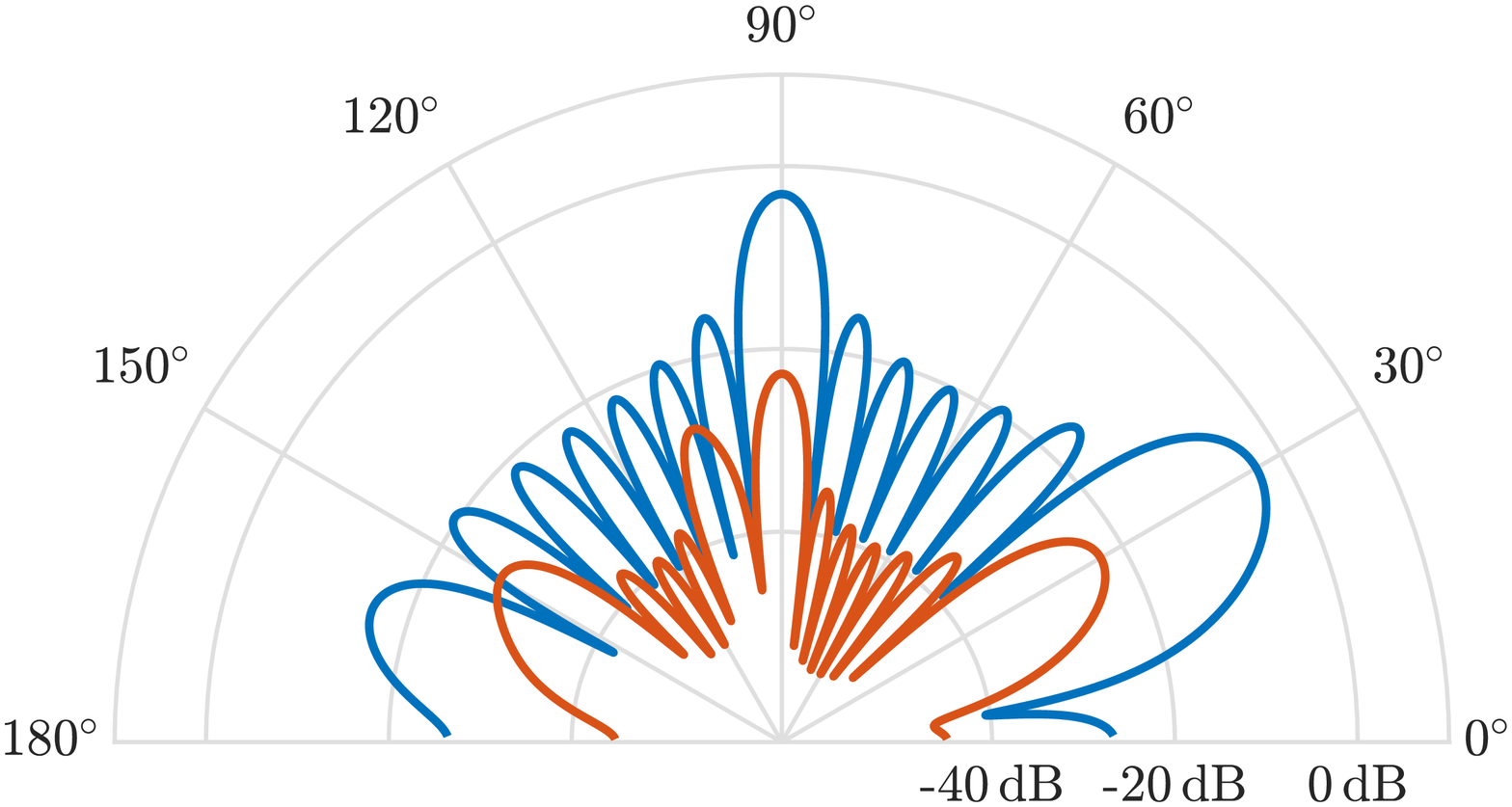}\label{fig:pattern_zf_2user}} \\
\subfloat[DAB; $K=1$ user, $\psi_{1} = 90^\circ$, and $\text{SNR} = -10$\,dB.]{\includegraphics[width =.44\textwidth]{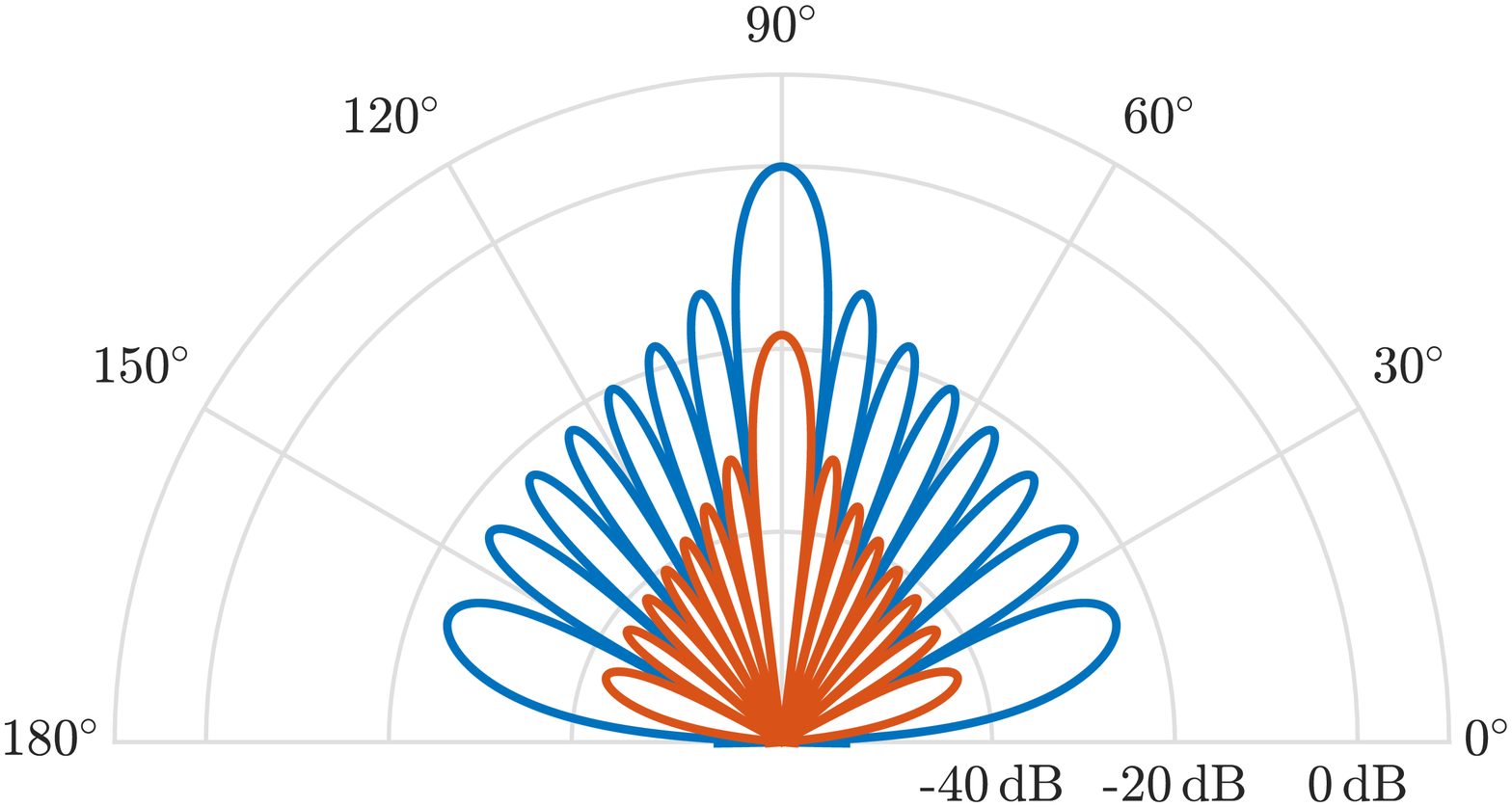}\label{fig:pattern_dab_1user_low}} \quad\quad
\subfloat[DAB; $K=2$ users, $\psi_{1} = 30^\circ$, $\psi_{2} = 90^\circ$, and $\text{SNR} = -10$\,dB.]{\includegraphics[width =.44\textwidth]{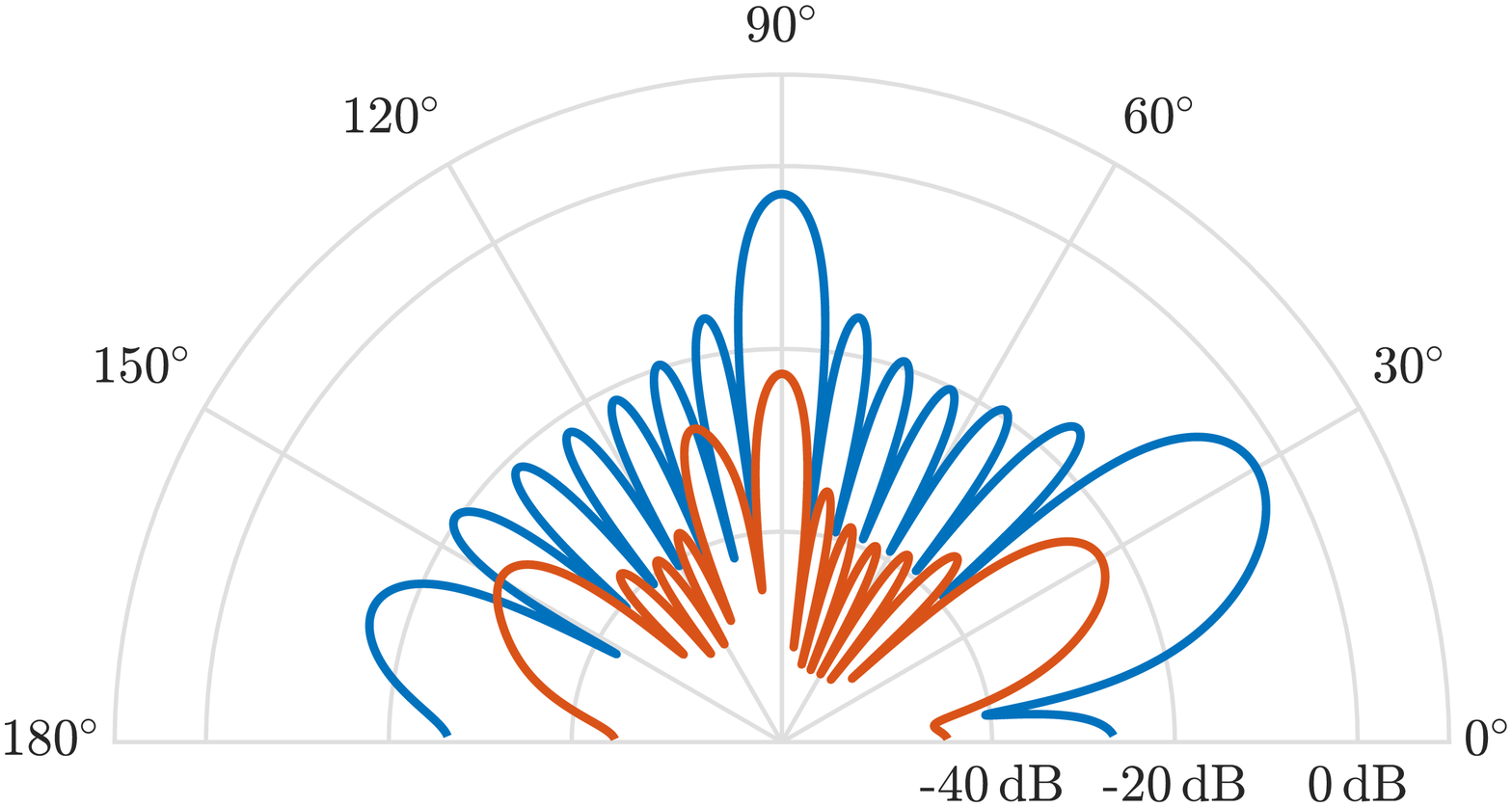}\label{fig:pattern_dab_2user_low}} \\
\subfloat[DAB; $K=1$ user, $\psi_{1} = 90^\circ$, and $\text{SNR} = 30$\,dB.]{\includegraphics[width =.44\textwidth]{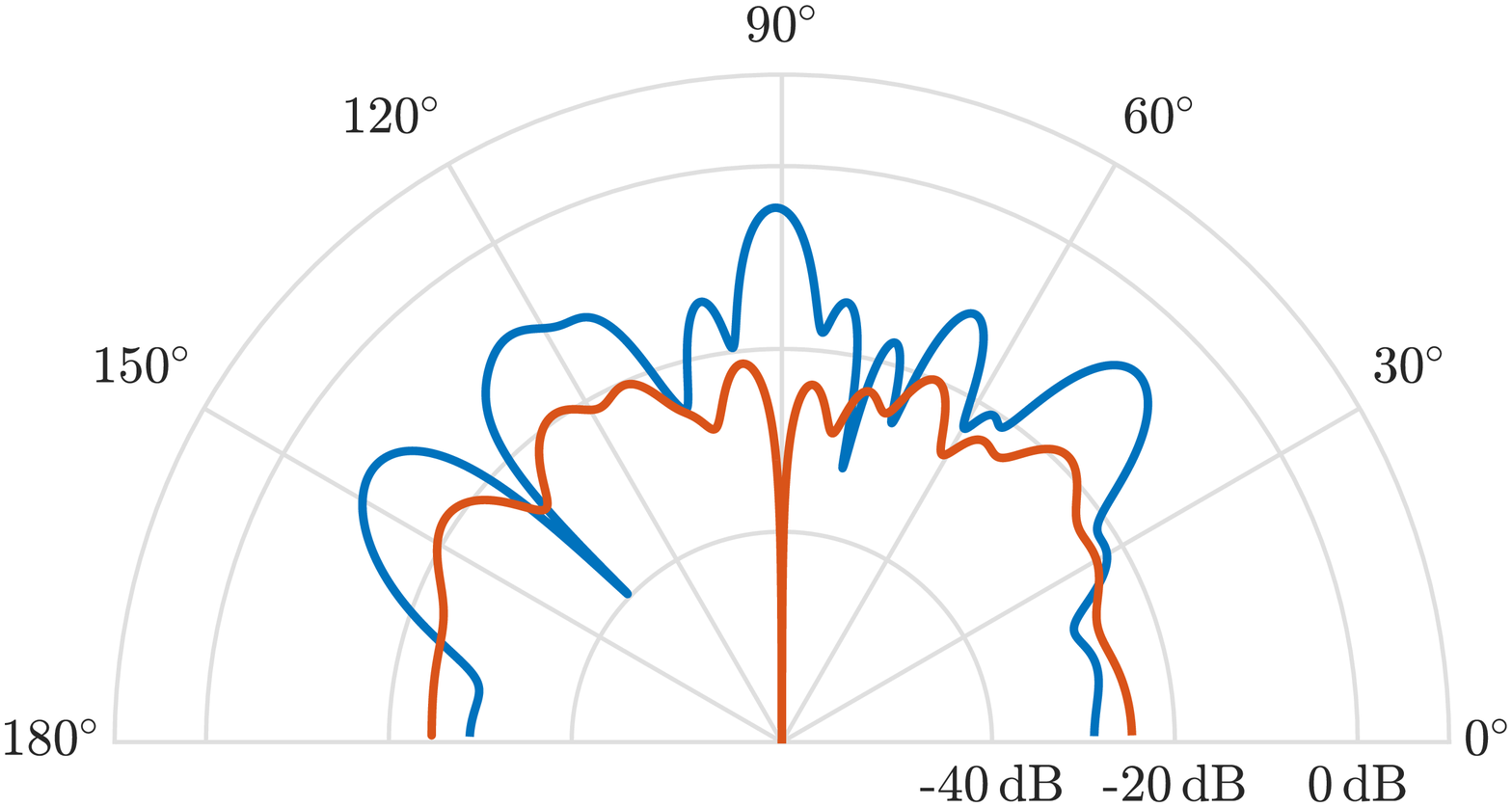}\label{fig:pattern_dab_1user_high}} \quad\quad
\subfloat[DAB; $K=2$ users, $\psi_{1} = 30^\circ$, $\psi_{2} = 90^\circ$, and $\text{SNR} = 30$\,dB.]{\includegraphics[width =.44\textwidth]{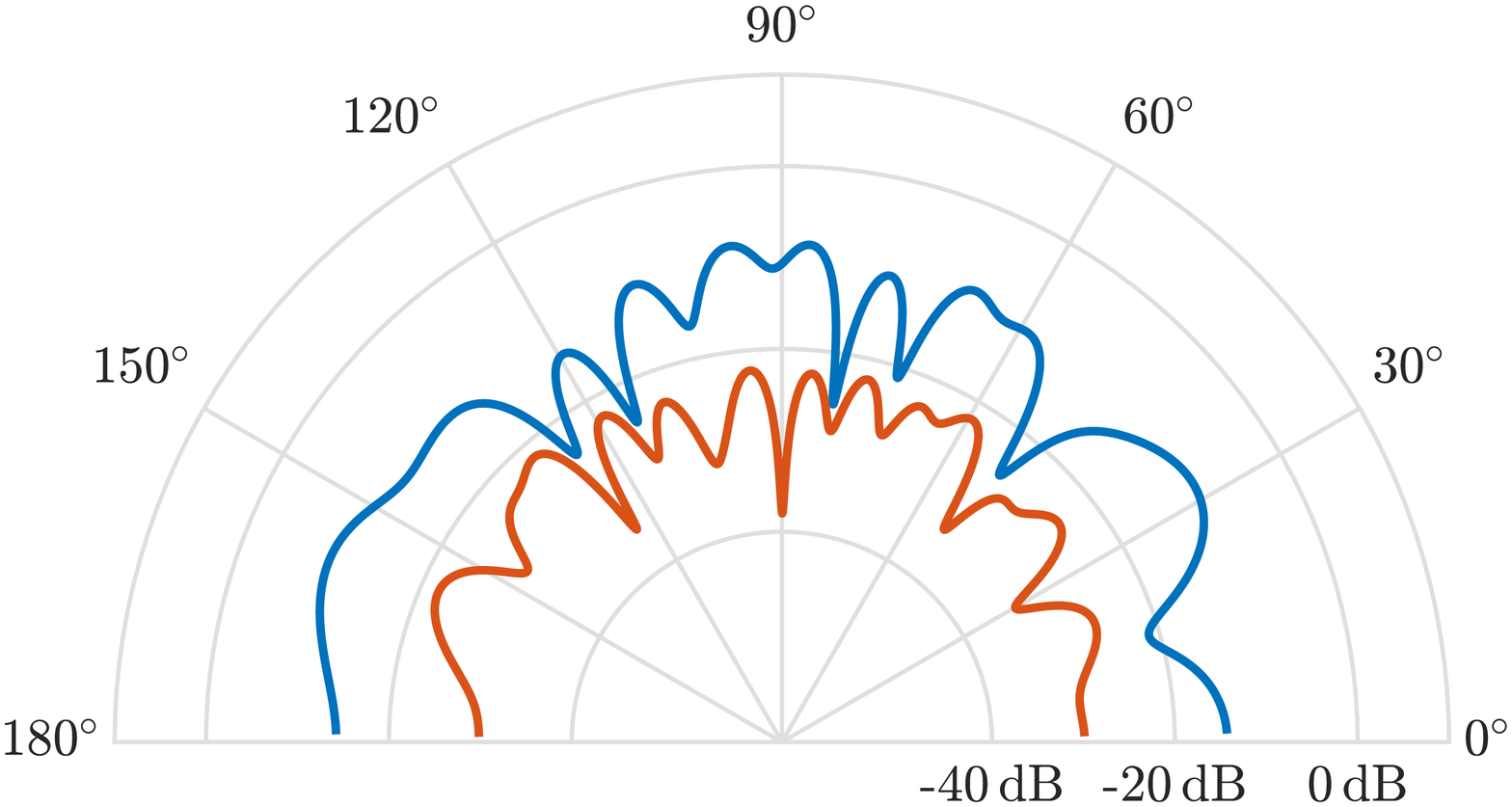}\label{fig:pattern_dab_2user_high}}
	\caption{Far-field radiation pattern for different linear precoders; $M = 16$ antennas, $K \in \{ 1, 2\}$ users, $\text{SNR} \in \{ -10, 30\}$\,dB, $\beta_1 = 0.98$, $\beta_3 = -0.04 - 0.01j$, and $P_\text{tot} = 43$\,dBm. The blue curves correspond to the linear term of the transmitted signal and the red curves correspond to the distortion~term. At high SNR, the proposed DAB precoder nulls the distortion in the direction of the users. At low SNR, the DAB precoder coincides with the MRT precoder.}
\label{fig:radiation_pattern}
\end{figure*} 

\subsection{Far-Field Radiation Pattern} \label{subsec_Perf}

To understand why DAB precoding outperforms conventional linear precoders at high SNR, we illustrate in \fref{fig:radiation_pattern} the far-field radiation pattern of the transmitted signal for $K \in \{ 1, 2\}$ users and $\text{SNR} \in \{ -10, 30\}$\,dB.
Recall from~\eqref{Bussgang} that the transmitted signal can be written as the sum of a linear function of the input to the nonlinearity and a distortion term.
The power of the linear component in the direction $\psi$ is $\veca^T(\psi) \matB \matP\matP^H \matB^H \veca^*(\psi)$; the power of the distortion term in the direction $\psi$ is $\veca^T(\psi) \matC_\vece \veca^*(\psi)$.
In~\fref{fig:radiation_pattern}, for the case $K = 1$, we set $\vech_1 = \veca(\psi_1)$, with $\psi_1 = 90^\circ$. For the case, $K = 2$, we set $\vech_1 = \veca(\psi_1)$ and $\vech_2 = \veca(\psi_2)$, with $\psi_1 = 30^\circ$ and $\psi_2 = 90^\circ$. For reference, we also show the far-field radiation pattern for MRT precoding, which is not dependent on the SNR operating point. 

We observe from \fref{fig:pattern_zf_1user} and \fref{fig:pattern_zf_2user} that with MRT precoding, the distortion is beamformed towards the users.
For high SNR values, DAB precoding, on the other hand, minimizes the distortion transmitted in the direction of the users in order to improve resulting SINDR. This becomes particularly evident in~\fref{fig:pattern_dab_1user_high}, where we observe that DAB precoding nulls the distortion in the direction of the user (recall that $\psi_1 = 90^\circ$ for this scenario) at the cost of reduced array gain to the users and increased radiation in unwanted directions.
This clarifies why DAB outperforms MRT and ZF in the high-SNR regime, where the performance is limited by the nonlinear distortion.
For low SNR values, where the performance is limited by the thermal noise, the DAB precoder instead maximizes the array gain in the direction of the users to maximize the received power. Hence, the DAB precoder coincides with the MRT solution (compare, e.g., \fref{fig:pattern_zf_1user} and \fref{fig:pattern_zf_2user} with~\fref{fig:pattern_dab_1user_low} and~\fref{fig:pattern_dab_2user_low}).

\section{Conclusions} \label{sec:Conc}

In this paper, we have proposed an iterative scheme for computing a distortion-aware linear precoder. The proposed scheme is shown to yield significant gains compared to conventional linear precoders over a mmWave multiuser MISO downlink channel for the case when nonlinear PAs are used at the transmitter.
We observed that, in the high-SNR regime and in the single-user case, the proposed algorithm is able to null the distortion in the direction of the user.

While the current work considers only a fully-digital precoding architecture, an extension to consider hybrid precoding is part of ongoing work. An extension of the proposed algorithm to other hardware impairments is also left for future~work.

\begin{spacing}{0.98}
\bibliographystyle{IEEEtran}
\bibliography{IEEEabrv,confs-jrnls,publishers,svenbib}
\end{spacing}

\end{document}